\documentclass[journal=jacsat,manuscript=article]{achemso}

\usepackage[version=3]{mhchem} 
\usepackage{xcolor}
\usepackage{verbatim}
\usepackage[british]{babel}

\author{Federica De Chiara}
\affiliation{School of Biomedical Engineering and Imaging Sciences, Faculty of Life Sciences and Medicine, King’s College London, London, SE17EU, UK}

\author{Hovan Lee}
\affiliation{Department of Physics, Faculty of Natural \& Mathematical Sciences, King's College London, London, WC2R2LS, UK}
\email{hovan.lee@kcl.ac.uk}

\author{Cedric Weber}
\affiliation{Department of Physics, Faculty of Natural \& Mathematical Sciences, King's College London, London, WC2R2LS, UK}

\author{Hongbin Liu}
\affiliation{School of Biomedical Engineering and Imaging Sciences, Faculty of Life Sciences and Medicine, King’s College London, London, SE17EU, UK}
\email{hongbin.liu@kcl.ac.uk}

\title[An \textsf{achemso} demo]
  {Sensing applied pressure by triggering \\ 
  electronic quantum many-body excitations \\ 
  in an optical waveguide}


\begin{document}







\begin{abstract}
Recently, nanomaterials are arousing increasing interest and a wide variety of optoelectronic devices have been developed, such as  light-emitting diodes, solar cells, and photodetectors.
However, the study of the light emission properties of quantum dots under pressure is still limited.
By using a joint theoretical and experimental approach, we developed a polymer waveguide doped with CdSe quantum dots for pressure sensing.
Absorption and re-emission effects of the quantum dots are affected by the pressure applied on the waveguide.
Specifically, since both amplitude and wavelength are modulated, not only the pressure can be detected, but also its location along the waveguide.
The calibration results demonstrate the feasibility of the proposed force sensor design.
Theoretical model and simulations further validate the presented sensing principle.
The proposed prototype benefits from the main advantages of optical sensors, such as their predisposition to miniaturization, small cable sizes and weights, immunity to electromagnetic interference, and safe operation in hazard environments.
In addition, bio-compatibility, non-toxicity and flexibility make the presented sensor potentially appealing to various application fields such as nanobiotechnology and robotic sensing.
\end{abstract}

\section{Introduction}
In the past few decades, nanotechnology has attracted a growing interest in various scientific and technological areas.
In particular, the development of novel nanomaterials is increasingly investigated.
Due to their size-related effects leading to unique physical, chemical, optical, electrical, and magnetic properties, these materials can be widely used to develop novel sensors \cite{NanomatSensors}. Among the different types of sensing technologies, optical based sensors offer important advantages: they enable small sizes, offer sensitivity to multiple environmental parameters, require small cable sizes and weights.

In contrast to electrical schemes, optical sensing offers potentials of high sensitivity, immunity to electromagnetic interference, and safe operation in explosive or combustive atmospheres, as well as more options for signal retrieval from optical intensity, spectrum, phase, and polarization \cite{OpticalSensors}. Nanomaterial-based optical sensors can also be expected to exhibit further advantages over conventional sensors.
In particular, the integration of nanoparticles in polymer hosts and their use in combination with optical fibers or planar platforms have attracted interest. 
Such arrangements are a key step towards the development of advanced analytical instrumentation, aiming small scale and multiparameter capability \cite{QDsInHost}.

The unique features of quantum dots (QDs) are highly attractive for fiber sensing platforms.
Polymer fibers are excellent candidates to host QDs because of the good processability of their organic materials, which facilitate the nanoparticle integration \cite{PolymerFibers}.
Polymer optical fibers and waveguides have received continuous attention in recent years due to their excellent properties, such as simple preparation, biocompatibility, low cost, high optical transparency and flexibility.
Although organic dyes were traditionally used for doping, they have been gradually superseded in several fields by quantum dots, because of their unique controlled size-dependent electronic and optical properties \cite{QDs_Dyes}.
In addition, the QDs brightness is enhanced by 20 times and the stability against photobleaching is improved by 100 times compared to organic dyes \cite{QDs_Dyes2}.\\
At present, some of the most attractive features of QDs are being explored for the detection of specific molecules in biomedical researches and clinical diagnosis \cite{Biosensors1,Biosensors2,Biosensors3,Biosensors4}.
Polymer embedded nanocrystals are also widely employed as luminescent indicators for detection of chemical species in sensing probes \cite{PolChemSensors1,PolChemSensors2}.
The unique features of QDs are also highly attractive for fiber sensing platforms. 
In spite of this, to date, the types of solutions where QDs are simultaneously immobilized and allowed to interact with the environment for sensing purposes are still very limited and the application of QDs in optical fiber technology remains largely unexplored. \cite{QDsInHost}.
One of the most investigated fields involves the use of QDs integrated in polymer optical fibers (POF) or planar waveguides for thermometry applications \cite{TempPOF1,TempPOF2,TempPOF3,TempPOF4,TempPOF5,WaveguideTempSensing1,POFsTempSensing}.
Among other relevant application fields, fluorescent fiber probes based on nanomaterials were employed to detect metal ions for monitoring of heavy metal pollution \cite{MetalIonsSensing1}, as well as for intracellular sensing and medical diagnostics \cite{MetalIonsSensing2}.\\
Nanomaterials provide a wide range of applications in nanopiezotronics, which utilizes the coupled piezoelectric and semiconducting property of nanomaterials.
Piezoelectricity is a coupling between mechanical and electrical behavior of a material.
When a piezoelectric material is squeezed, twisted, or bent, the transport behavior of electric charges is largely influenced.
Conversely, when a piezoelectric material is subjected to a voltage drop, it mechanically deforms.
Electronics based on this effect are coined as piezotronics \cite{Piezophototronics0}.
In recent years, a large variety of piezotronic devices have been invented, such as strain sensors, transistors, logic circuits, and electromechanical memories \cite{Piezophototronics1}.
By introducing light illumination, piezoelectric properties affect the photoluminescence emission of nanomaterials. 
This effect leads to the development of a new research area named piezophototronics \cite{Piezophototronics2}.
At present, the piezophototronic effect is extensively applied to enhance the performance of a variety of optoelectronic devices, such as light-emitting diodes, solar cells, and photodetectors \cite{Piezophototronics1}.
The study of the piezophototronic effect of QDs integrated in polymer waveguide and optical fibers can open to interesting possibilities and lead to the development of competitive devices.

In this paper, the core/shell CdSe/ZnS QDs were integrated in a flexible millimeter-scale polymer waveguide.
When a force is applied on the waveguide, the light emission detected at the output changes as a result of the piezophototronic effect of the QDs contained into the waveguide core.
A calibration experiment of intensity modulation and wavelength shift as a function of the force applied on the waveguide demonstrates the feasibility of the proposed force sensor design.
Theoretical model and simulation results further validate the presented sensing principle.

\section{Results and discussion}

\subsection{Materials selection and manufacturing process}

Optical fibres are made by an inner cylindrical core and an outer covering layer called cladding.
They are able to transmit light because of the phenomenon of \textit{total internal reflection} (\textit{TIR}).
Snell's law well describes the behaviour of the light at the interface between two media. 
From Snell's law, the equation of the critical angle can be deduced as follows:
$$
\theta_{c} = \arcsin(\dfrac{n_{2}}{n_{1}})
$$
\noindent
Assuming that the refractive index of the first medium $n_{1}$ is higher than the refractive index of the second medium $n_{2}$ and calling $\theta$ the angle between the propagation direction of a light ray and the axis perpendicular to the interface between the two media, all the rays having $\theta > \theta_{c}$ are totally reflected into the first medium.
In the case of an optical fibre, if $n_{1} > n_{2}$ with $n_{1}$ refractive index of the core and $n_{2}$ refractive index of the cladding, all the rays satisfying the above angular condition are transmitted through the core.\\
Some PTFE sleeves (clear Polytetrafluoroethylene, 3.05mm bore diameter, RS components) were employed as cladding.
The main material being used to create the core was PMMA powder (Polymethyl methacrylate, average Mw ~120,000 by GPC, Sigma-Aldrich).
The typical refractive index value is around 1.50 at 500 nm for PMMA and 1.38 at 500 nm for PTFE. Since a low concentration of quantum dots was embedded into the core of the optical fibre (as detailed below), it can be assumed that the nanoparticle integration does not affect the refractive index of the core, therefore $\theta_{1} > \theta_{2}$ as required for TIR to occur.

The PMMA powder was diluted in the organic solvent DMF (N,N-Dimethylformamide, anhydrous, 99.8\%, Sigma-Aldrich) at 22wt\%. The solution was mixed with a stirrer at 1100 rpm for about 5 hours.
The core-shell CdSe/ZnS quantum dots with emission wavelength of 560 nm (powder form, PlasmaChem GmbH) was diluted in toluene (Anhydrous, 99.8\%, Sigma-Aldrich) with a mass concentration of 5mg ml$^{-1}$.
Quantum dots were added to the polymer solution in the concentration of 0.06wt\% and mixed for 30min using the stirrer. 
The final solution was placed in the ultrasonic bath for 15 minutes to remove air bubbles.
Then the polymeric composite material was injected into the PTFE tubes using a syringe.
The filled tubes were left to dry for 24 hours at room temperature and then for 72 hours at 60°C in the oven.
After the curing process, the polymeric composite material containing quantum dots has a gel-form and this allows the optical fibre to be flexible.\\
The fabrication process of the plain optical fibre (not containing quantum dots) follows the same procedure as described above, except the phase of quantum dots integration.

\begin{figure}[t]
    \centering
    \includegraphics[width=\linewidth]{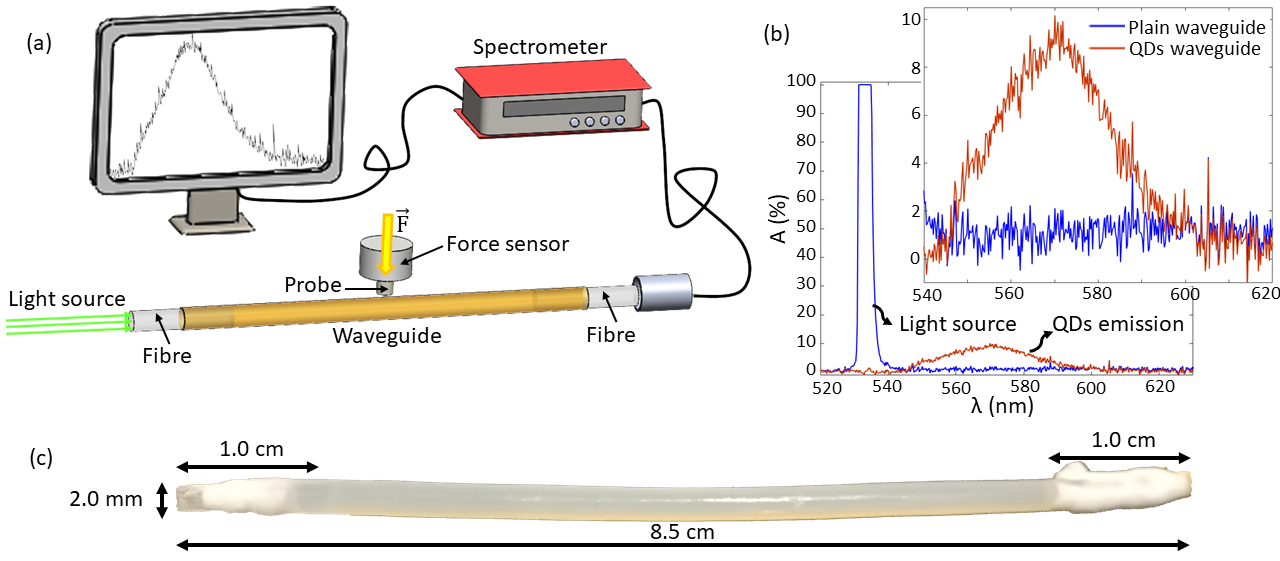}
    \caption{\textbf{Quantum dots piezoelectric effect for sensing in composite polymer waveguides.}
    a) Experimental setup: a collimated laser diode with emission peak around 532nm injects light into the polymer waveguide. The output signal is collected by an optical system and sent to the spectrometer which is connected to a PC. A pressure is applied by a probe mounted on a force sensor.
    b) The output spectrum was acquired using a plain polymer waveguide (without QDs) and a polymer waveguide with QDs incorporated in its volume. 
    The zoomed plot highlights the range [540-620]nm: considering the plain waveguide, the spectrum appears flat, while the signal corresponding to the QDs piezoelectric emission stretches on the range [550-600]nm in the case of the doped waveguide.
    c) The main materials for the polymer waveguide fabrication are Teflon (PTFE) for the cladding and Plexiglass (PMMA) in the form of a gel for the core. 
    The whole waveguide is flexible and compressible.
    }
    \label{fig:setup}
\end{figure}

\subsection{Experimental setup}

The experimental setup is shown in Fig.\ref{fig:setup}a.
A collimated laser diode ($\lambda_{em} \approx$ 532nm, 4.5mW, Thorlabs) injects light into a cylindrical polymer waveguide of 8.5cm length and diameter of 2mm.
A long-pass filter ($\lambda_{cut-on} \approx$ 550nm, Thorlabs) was placed at the output of the waveguide to filter out the light coming from the laser and isolate the signal emitted by the quantum dots.
The output signal is collected by an optical system and sent to the spectrometer (range [200-1000]nm, Thorlabs). 
The acquired spectrum is displayed through a software on a PC.
Fig.\ref{fig:setup}b shows a comparison between the spectrum acquired using a plain waveguide (not containing quantum dots) and the one employing a waveguide with embedded quantum dots.
For both the acquisitions, the same integration time was set on the spectrometer.
In the case of the plain waveguide some laser light can still reaches the spectrometer although the presence of the long-pass filter and its peak saturates around 532nm, while the rest of the spectral profile is flat (blue spectrum in fig. \ref{fig:setup}b).
In the doped waveguide instead, nanoparticles absorb part of the laser light and eventually re-emit it at longer wavelength.
So, a spectral distribution stretches on the range [550-600]nm (red spectrum in fig. \ref{fig:setup}b).
This emission corresponds to the quantum dots piezoelectric emission and it can be detected also when the waveguide is not compressed.
In fact, even under resting conditions, the quantum dots are affected by an inner strain for being integrated in the gel-like matrix of the waveguide core and because of the injection process undergone by the soft material during manufacturing.

The proposed waveguide is shown in Fig.\ref{fig:setup}c.
An optical fibre of 1 cm length was inserted at the input and output of the PTFE tube corresponding to the waveguide cladding.
A fibre with core diameter similar to the waveguide core diameter (3mm) was employed to reduce the light loss at the interface between the optical fiber and the soft material of the waveguide core. 
The optical fibres were firmly fixed inside the PTFE tube using some thread sealing tape.
When a force is applied on the waveguide, the two optical fibres assume the role of stoppers and keep the soft material of the waveguide under compression.
The pressure is then sensed by the quantum dots embedded into the gel-like matrix and their light emission changes as a result of the piezophototronic effect.

\begin{figure*}[t]
    \centering
    \includegraphics[width=\linewidth]{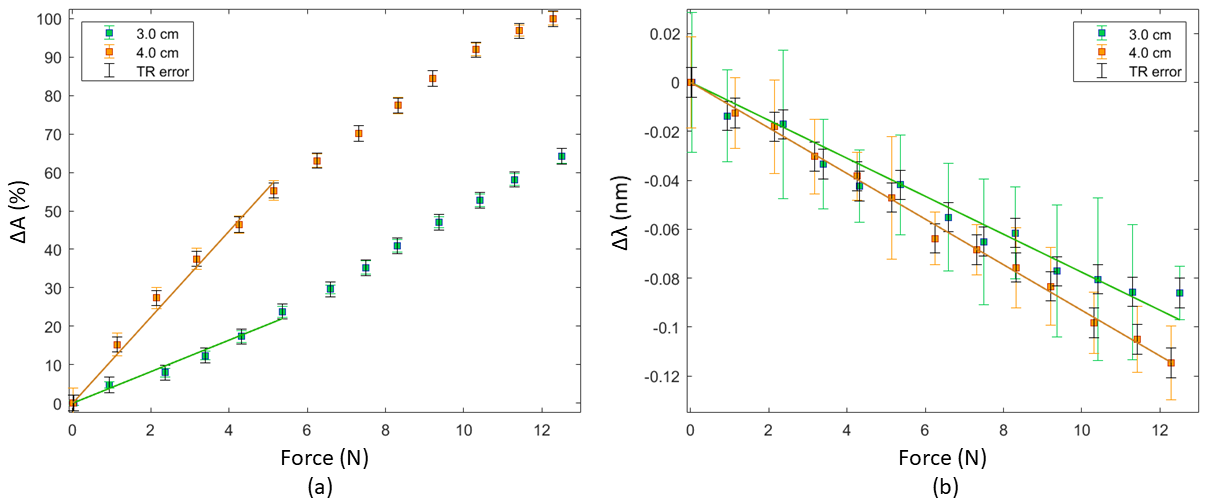}
    \caption{
    \textbf{Amplitude modulation and wavelength shift calibration.}
    The waveguide was compressed on two different locations: 3 cm and 4 cm from the input surface of the waveguide.
    For each force value, the amplitude of the piezoelectric peak of quantum dots a) and its wavelength shift b) were measured.
    }
    \label{fig:Calibration}
\end{figure*}

\subsection{Experimental evaluation}

A calibration experiment was carried out to determine the characteristic trend of intensity and wavelength modulation as a function of the force applied on the waveguide.\\
A 6-axes force/torque sensor (ATI Nano-17, resolution 0.003 N) was mounted on a manual translation stage (Thorlabs, resolution 10 $\mu$m) which was moved forward to progressively compress the waveguide (fig.\ref{fig:setup}a).
A circular probe of 8 mm diameter mounted on the calibration force sensor was employed to apply the compression on two different locations: 3cm and 4cm from the input surface of the waveguide.
The force applied covers the range [0-12]N, with a step of 1N. 
The maximum value of 12N is the mechanical limit imposed by the experimental setup in use and the amplitude corresponding to this force value was used as normalisation factor.
The integration time of the spectrometer was set to 5s to get a good signal-to-noise ratio. 
For each force value, 10 repeated spectra were acquired.
A Gaussian fit was applied on each one of the 10 spectra, then the mean of the coefficients was calculated to get the final Gaussian fit for each force value.
The amplitude coefficient of the final Gaussian trend was considered to get the modulation of the peak amplitude of the signal (fig.\ref{fig:Calibration}a), while the centroid location coefficient was used to get the wavelength shift as a function of force (fig.\ref{fig:Calibration}b).
The error bars were calculated as standard deviation on the 10 repeated measurements for each force value. 
The black error bars take into account the stress relaxation occurring during the acquisition time for each force value. 
This effect will be discussed in the next paragraph.

\subsection{Stress relaxation}

\begin{figure*}[t]
    \centering
    \includegraphics[width=\linewidth]{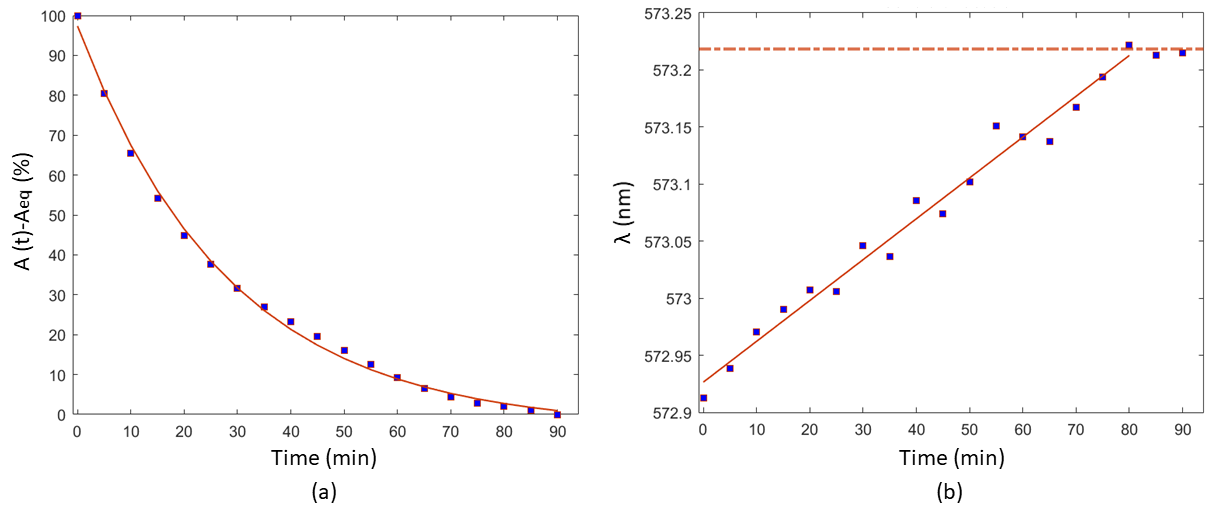}
    \caption{
    \textbf{Time-dependent relaxation of the polymer waveguide.}
    The waveguide was kept under constant pressure for 90 minutes and the output spectum was acquired every 5 minutes.
    a) The amplitude modulation as a function of time $A(t)$ shows a negative exponential behaviour, typical of stress-relaxation in viscoelastic materials, especially polymeric materials.
    The amplitude $A(t)$ reaches its equilibrium value $A_{eq}$ after 90 minutes of acquisition. 
    b) The wavelength shift presents a linear fashion.
    Both the trends move towards a plateau after about one hour and an half of acquisition.
    }
    \label{fig:StressRelax}
\end{figure*}

Polymer materials show a combination of elastic and viscous behaviour known as viscoelasticity.
An immediate consequence of viscoelasticity is that deformations under stress are time dependent.
If a constant deformation is imposed then the induced stress will relax with time (stress relaxation).
Several polymeric materials show a linear viscoelastic behaviour, such that the stress (at a constant strain) decreases in time as a simple exponential Debye function:
\begin{equation}
\sigma(t)=\sigma_0 exp^{-t/\tau}
\end{equation}
The time constant $\tau$ is the relaxation time at which $\sigma(t)$ decays to the value $1/e$ of
its initial value $\sigma_0$\cite{StressRelax}.\\
To study the time dependency of amplitude and wavelength of the output signal, a constant pressure was applied on the waveguide and the spectrum was monitored for 90 minutes.
Results are shown in Fig.\ref{fig:StressRelax}.
The exponential decay of the amplitude over the time suggests that the polymeric soft material of the waveguide core is subject to a stress relaxation effect.
Both amplitude and wavelength move towards a plateau after about one hour and an half of acquisition. 

\subsection{Analysis of simulation results}

\begin{figure}
    \centering
    \includegraphics[width=\linewidth]{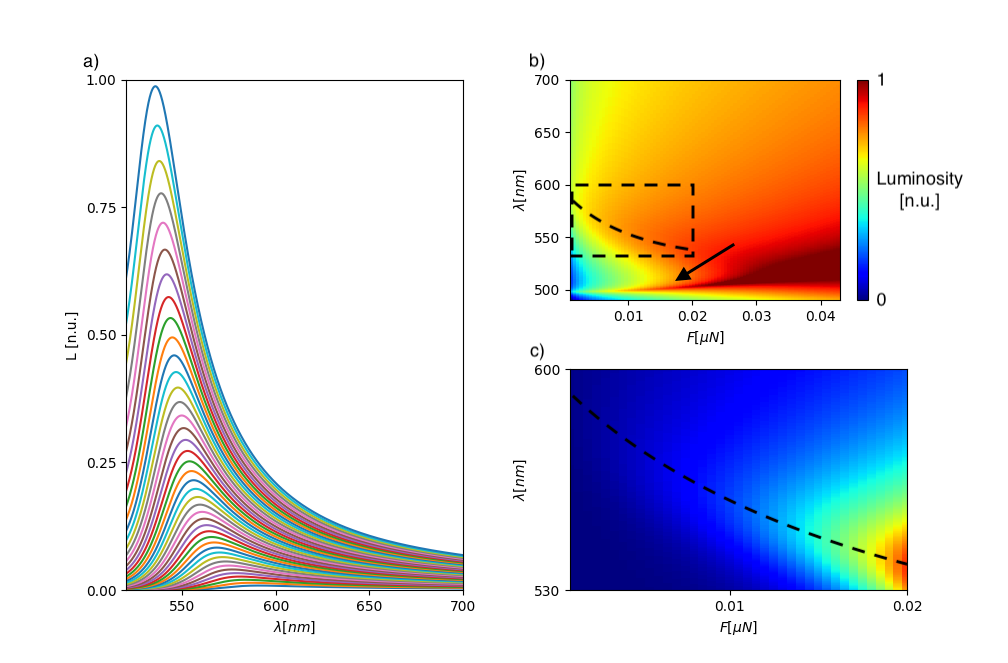}
    \caption{\textbf{Single quantum dot calculated photoemission spectra.}
    a) Normalised simulated photonic emission luminosity spectra of a single quantum dot under compression, shown here for a range of compression force values. We note that the trend of the feature at the 540 to 580 $nm$ range blue shifts and increases in amplitude under compression.
    b) Heat map of the single quantum dot photoemission effect. The range of 
    compression is evaluated from the force of 0 to 0.04 $\mu N$.
    The lower wavelength band is associated with the QD effective energy, whereas the upper wavelength band corresponds to the
    piezoelectric dipole emission (see text). Upon the application of stress, the excitonic band is slightly
    red shifted with pressure, whereas the piezo-electric driven emissions are blue shifted. Dotted-line box indicates the region shown in c). We note that
    the emission is simulated for a single QD in vacuum. 
    c) Blown up scales to highlight the regime of forces
    which is comparable to experiments. Dotted line as a guide to the eye for the blue shift trend of the piezo-electric feature under compression. The colour scheme for b) is logarithmic whereas c) is linear.}
    \label{single}
\end{figure}

The normalised single quantum dot emission spectra are illustrated in Fig.\ref{single}.a) where the increase in emission amplitude and the blue shift in wavelength, as the compressive force increases, for the piezoelectric band is shown.

The full single quantum dot emission spectrum under compressive forces of range $0-0.04 nN$ is shown as a heat map in Fig.\ref{single}.b), here we observe two distinct features: the quantum dot effective energy peak at shorter wavelengths and the piezoelectric dipole emissions for longer wavelengths (as clarified in the Theory section). The two features undergo contrasting wavelength shifts under an increase in force; the shorter wavelength band redshifts, whereas the longer wavelength band blueshifts. This presents an upper limit of the operational force range of this technique due to the adjoining of the two features at $\sim 0.03 nN$, at least at the single quantum dot emission spectra level. 

In Fig.\ref{single}.c) we showcase the single quantum dot operational force range
(the same region is shown in Fig.\ref{single}.b) in a dotted box). The blue shifting trend is explicitly depicted as a dotted line.

To further justify our experimental observations, the single quantum dot emission spectra were utilised in the ray tracing simulation; the waveguide core was split into three separate simulation segments of refractive index $n_{PMMA}=1.49$, dimensions $50\times50\times50 nm$ each, surrounded by a cladding along the length of the waveguide of refractive index $n_{PTFE}=1.38$, thickness $10 nm$. The rays are generated in a spherical cone distribution of $30 rad$ at the beginning of the first segment, which also contains spherical structures of refractive index\cite{cdse_n} $n_{CdSe}=2.64$ (radius = $1.5nm$, consistent with the manufacturer values) distributed in a cubic mesh manner with inter-quantum dot distance of $10 nm$
. These spherical structures play the role of the quantum dots suspended in the waveguide, and contain absorption and emission spectra corresponding to the ambient force (as calibrated through matching the simulation results with experimental observations) within the PMMA. The second and third segments contains simulated quantum dots with different spectra, consistent with a further externally applied force.

The ray tracing simulation takes several types of events into account as a ray approaches any interface between objects; the ray could be reflected and adopt a new direction of propagation, or transmitted at angle due to the difference of refractive indices at the interface. The ray could also be absorbed, corresponding to the absorption spectrum of a quantum dot. Finally after being absorbed into a quantum dot, the ray has a likelihood be being emitted from a quantum dot with a new wavelength and propagation direction.

\begin{figure}
    \centering
    \includegraphics[width=\linewidth]{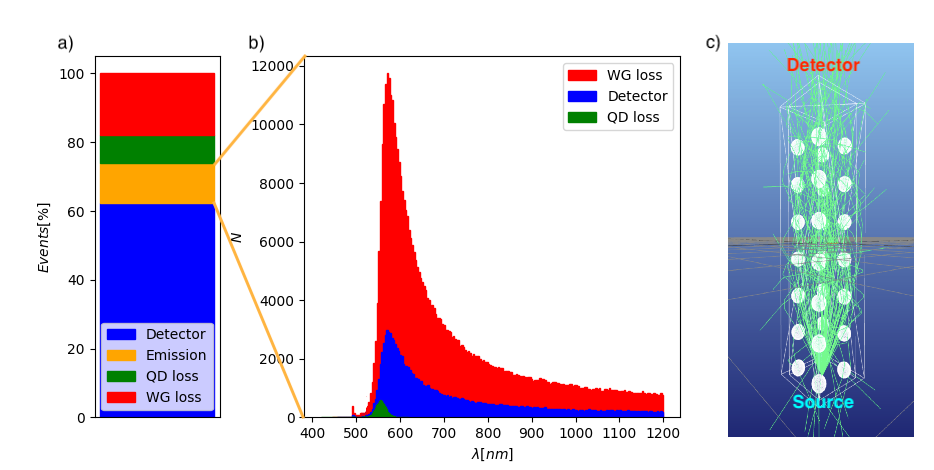}
    \caption{\textbf{Ray-tracing outcomes.}
    a) Stacked bar chart illustrating the breakdown of ray tracing events, where: i) 'Detector' indicates that a ray reaches the detector without being absorbed by and/or emitted from a QD, ii) 'Emission' denotes the subset of rays that is absorbed and emitted by a QD at least once, iii) 'QD loss' depicts rays that are absorbed by a quantum dot, and due to the quantum efficiency of the quantum dot, are not emitted again, iv) 'WG loss' (waveguide loss) accounts for rays that escapes the waveguide through the outer coating of the waveguide. b) Stacked histograms resolved in wavelengths showing the further split of the 'Emission' events in a). c) Wireframe structure diagram of a scaled down example of the ray tracing simulation.}
    \label{events}
\end{figure}

The percentages of final outcome of the rays for a sample ray tracing simulation of five million rays, at ambient waveguide conditions, are shown in Fig.\ref{events}.a). The outcomes are split into four categories; all rays that have not been absorbed or emitted by a quantum dot, and are transmitted through the detector-end interface between the simulated waveguide core and vacuum are shown as a part of the 'Detector' category in blue. Rays that have not been absorbed or emitted by a quantum dot, and are not transmitted through the waveguide core/ detector interface are depicted in the 'WG loss' category in red, indicating that the rays have been lost due to transmission through the waveguide cladding into simulated vacuum. The rays that have been absorbed by a quantum dot, and are not re-emitted due to the quantum yield of the quantum dots (estimated at $60\%$) are shown in green as 'QD loss'. Lastly, all rays which have been emitted by a quantum dot are classified in the 'Emission' category in orange.  

The 'Emission' rays are further split into three wavelength resolved outcome categories in Fig.\ref{events}.b), where the classification of ray categories follow those of Fig.\ref{events}.a), but all rays have been emitted by a quantum dot.

A comparison between the ratio of 'Detector' rays in Fig.\ref{events}.a) and 'Detector' rays in Fig.\ref{events}.b) informs us that we are in the regime where the detector senses an order of magnitude more light-source wavelength rays than quantum dot emitted (or signal) rays, consistent with the experimentally observed ratios where the light-source wavelength is detected as a saturated peak. However, due to the lack of information on the thickness of quantum dot shells, we were unable to calculate the relative densities of PMMA and quantum dots. Therefore, we could not use the simulation results to guide the manufacturing process, in order to experimentally obtain the optimal signal to light-source ratio.

Moreover, a significant amount of simulated rays are lost through the 'WG loss' categories. This is due to ray being scattered or emitted by quantum dots at angles with large perpendicular components with respect to the length of the waveguide. This allows for the possibility where rays propagate to the cladding at an angle greater than $\theta_c$, whereby the ray is transmitted through the cladding and are lost. This phenomena also explains the significantly larger ratio of 'WG loss' in Fig.\ref{events}.b), where the rays have been emitted by quantum dots, compared to 'WG loss' in Fig.\ref{events}.a), where the rays have not been emitted by quantum dots and are likely to propagate with a large parallel component with respect tot the length of the waveguide. 

\begin{figure}
    \centering
    \includegraphics[width=\linewidth]{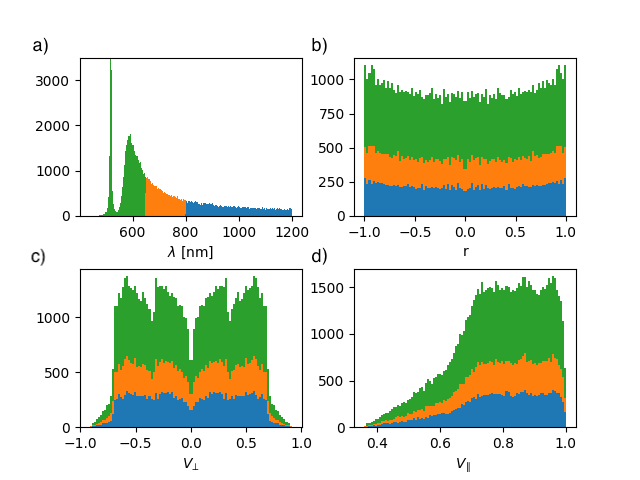}
    \caption{\textbf{Statistical analysis of simulated detected photons. }
    a) Spectral analysis of the detected photons after transmission through the waveguide with a simulated compressive force of $6 nN$.
    The subset of measured photons are colour coded into three groups: $\lambda <= 650nm$, $650nm < \lambda <= 800nm$ and $800nm < \lambda$. b) Spatially resolved cross-section of the measured photons. c) We also report the 
    angular distributions in terms of normalised momentum vector with respect to the c) perpendicular and d) parallel components.}
    \label{hist}
\end{figure}

The 'Detector' category in Fig.\ref{events}.b) can be further analysed to obtain the distribution of ray location and propagation directions. For a simulated compressive force of $6 nN$ on the quantum dots of the middle waveguide segment, all rays (of a five million ray simulation) that have been emitted by a quantum dot and is transmitted through the detector end of the waveguide is shown in Fig.\ref{hist}.a). Here, the rays are colour coded into three groups ($\lambda <= 650nm$ for green, $650nm < \lambda <= 800nm$ for orange and $800nm < \lambda$ for blue). The position of the ray incident on the simulated waveguide-detector interface is illustrated in Fig.\ref{hist}.b), where r is the normalised uni-dimensional distance of the ray position from the centre of the cross-section of the interface. The normalised component of ray propagation which is perpendicular with respect to the length of the waveguide is shown in Fig.\ref{hist}.c), and the parallel component is shown in Fig.\ref{hist}.d).

We observe in Fig.\ref{hist}.b) that there is a slight preference for rays to propagate incident to the interface with r close to unity. In our view, this is due to rays that are consistently reflected between the waveguide core-cladding interface and the outermost quantum dots from the centre of the waveguide. 

In fig.\ref{hist}.c) there is a tendency for rays to transmit through the interface with a nonzero perpendicular component. We explain this bias by considering rays with purely zero perpendicular component, since Fig.\ref{hist} show only the subset of rays that have been emitted by a quantum dot, rays that have been emitted with zero perpendicular component (and hence travel parallel to the length of the waveguide), will likely interact with the next neighbouring quantum dot in the cubic mesh. Therefore, this decreases the likelihood that the detector senses a ray with a zero perpendicular component. \textcolor{black}{Furthermore, we observe a disinclination of rays with a normalised perpendicular component of $\sim 0.35$. We attribute this to the scattering of rays in the quantum dot mesh}. Lastly, rays with unity perpendicular components will not propagate to the interface, and hence we observe zero counts of such rays.

In fig.\ref{hist}.d) a propensity of rays with $0.7 - 1$ normalised parallel components are observed. This can be explained through the geometry of the simulated waveguide; rays that transmits through the waveguide-detector interface likely has a large parallel component, such that no obstructing quantum dots are in the path of propagation. Furthermore, no counts of rays with a parallel component $< \sim0.35$ was detected. This is due to the critical angle of the interface; the angle of incidence of such rays are less than $\theta_c$ between the PMMA waveguide core and vacuum.

\begin{figure}
    \centering
    \includegraphics[width=\linewidth]{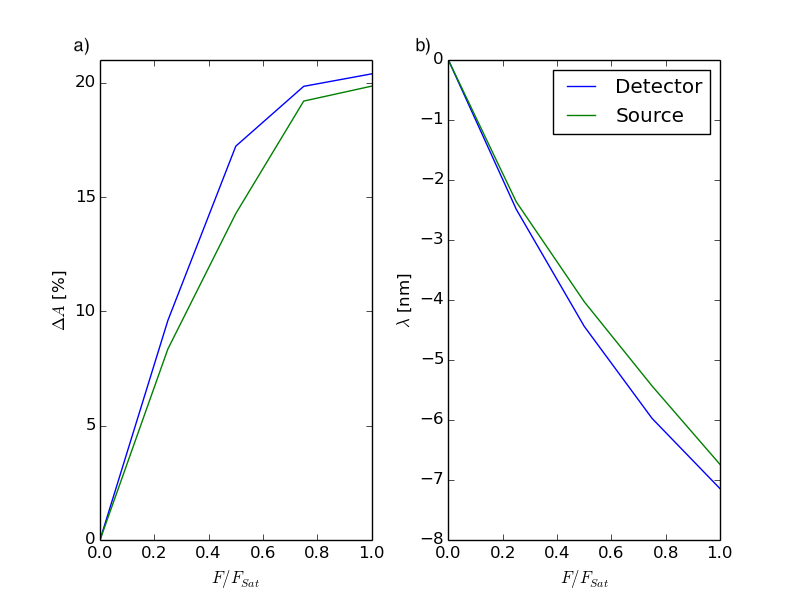}
    \caption{\textbf{Simulated trend of the intensity and wavelength.}
    a) Theoretical amplitude of the measured photonic distribution measured at the detector as a function of
    the applied force, where the latter is applied near the detector (source) and labelled in blue (green). 
    The distribution amplitude is obtained by a skewed Gaussian fit of the waveguide spectrum, evaluated as the percentage increase of the amplitude with respect to ambiant conditions ($\Delta A$) as a function of waveguide compression. The applied forces ($F$) is normalised by the saturation limit ($F_{sat}$). b) We report the wavelength shift of the maxima of the distribution's peak, as a function of waveguide compression.}
    \label{trend}
\end{figure}

We present the skewed Gaussian-fitted trend of simulated signal intensity as a function of compressive force applied onto the middle waveguide segment (green) and last waveguide segment (blue) in Fig.\ref{trend}.a), and the wavelength shift trend in Fig.\ref{trend}.b). The range of forces considered are normalised to a saturation force $F_{sat}$, where the amplitude increase of the signal saturates. This increase in amplitude and subsequent saturation is due to the growth and assimilation of the two features shown in Fig.\ref{single}.b); as the compressive force increases, the two features increase in amplitude and become closer in wavelength, whereby the fit interpret the two features as a single entity. That is, until the two features diverge in wavelength, and the fit only analyses the initially longer (now shorter) wavelength peak.

Both the intensity and wavelength trends of the 'Detector' and 'Source' simulations agree with the experimental results. Moreover, the trends are more pronounced for the 'Detector' simulations. We explain this due to the likelihood of a quantum dot emitted ray to become scattered out of the waveguide (WG loss), or be absorbed (QD loss) and re-emitted by a subsequent quantum dot; the closer to the detector this compressive force is applied, the less likely that the signal information is overwritten by these effects from a subsequent quantum dot.

\section{Theory}


In this section, we discuss the evaluation of the set of piezoelectric affected emission spectra from a single quantum dot under a range of compressive forces. A few difficulties exist in simulating these spectra; the crystalline structure and thickness of the ZnS shell are not specified by the provider. 

ZnS exist in both sphalerite (cubic) and wurtzite (hexagonal) forms, the piezoelectric tensor of the sphalerite form only contain a single unique nonzero element: $d_{14}$ (in Voigt notation), which does not correspond to a uniaxial compressive stress, and therefore can be safely neglected for our calculations. The relevant piezoelectric tensor elements of the wurtzite form are roughly a factor of three smaller than the corresponding CdSe tensor elements ($d^{ZnS}_{31}=-1.1$, $d^{ZnS}_{33}=3.2$, $d^{CdSe}_{31}=-3.92$, $d^{CdSe}_{33}=+7.8$, all values are in $pC/N$ and are cited from \cite{cdse_coeff}), and thus do not play a significant role in affecting the piezoelectric trend that we wish to showcase as a proof of concept. Moreover, since the ZnS shell of the quantum dots do not drastically alter the spectral line shape\cite{qd-shell}, we modelled our quantum dots as pure CdSe entities following the quantum mechanical analysis of piezoelectric effects from Zhang et al\cite{Zhang2018}.

In the cited work, rate of emission is calculated via the Lorentzian expression of the zero dimensional density of states and a product of Fermi-Dirac distributions:
\begin{equation}
    R(\omega)=\frac{\Gamma/2}{(\hbar\omega-E_{QD})^2+(\Gamma/2)^2}f_c(1-f_v),
\end{equation}
where $\Gamma$ is the line width (or full width at half maximum) of the distribution, which describes the coupling between a photon and an electron in the conduction band of the quantum dot. $f_c$ and $f_v$ are the Fermi-Dirac distributions of electrons in the conduction and valence bands of the quantum dot, and $E_{QD}$ is the effective energy of the quantum dot, described as:
\begin{equation}
    E_{QD}(r,F,T)=E_g+E_e(r)+E_h(r)-E_{ex}-E_p(r,F)+\alpha_1T
\end{equation}
with $E_g$ as the band gap of bulk CdSe, $E_{e,h}$ are the electron and hole confinement energies, and are inverse-squarely proportional to the radius of the quantum dot. $E_{ex}$ is the excitonic binding energy, for CdSe the electronic thermal energy (at room temperature and in vivo) exceeds the binding energy of excitons. Therefore, we neglect the excitonic binding energy. $E_{p}$ is the energy change induced by the piezoelectric potential of the material, it is dependent on both the radius of the quantum dot and the force applied onto the quantum dot. In addition, we have also included the shift in $E_{QD}$ due to temperature; the Debye temperature of CdSe ($\Theta_D=181K$\cite{cdse_coeff}) is significantly lower than room temperature, we therefore implemented a linear shift in energy with the trend $\alpha_1=0.32MeV/K$\cite{temperature}.

The calculation of the Fermi-Dirac distributions $f_{c,v}$ require the notion of chemical potential $\mu_{c,v}$ in both the conduction and valence bands, which are evaluated by considering the radius and force dependent, piezoelectrically generated charge carriers in the respective bands:
\begin{equation}
    n_{c,v}(r,F)=\frac{dF}{V}=\frac{1}{2\pi^2}(\frac{2m_r}{\hbar^2})^{3/2}\int\frac{\sqrt{E}}{e^{(E-\mu_{c,v})/k_BT}+1}dE
\end{equation}
where the charge carrier density is determined from the piezoelectric strain coefficient of CdSe $d$, the applied force $F$ and the volume of the quantum dot $V$, this can be further expressed through the notion of reduced mass $m_r$, with an integral over energy $E$.

The resultant spectra contains two identifiable features: the Lorentzian peak corresponding to the effective quantum dot energy, and the overlap between the piezoelectric charge Fermi-Dirac distributions, as illustrated in Fig.\ref{single}. The two structures are dependent on both the radius of the quantum dot and the force applied. However, the effective quantum dot energy peak red-shifts under increasing force, whereas the piezoelectric band blue-shifts. Moreover, the pair of structures depend differently on the radius of quantum dots. 

This therefore permits a rough calibration of the unique parameters which best describe the experimental setup, assuming that the experimental spectra of quantum dots suspended in a waveguide are indicative of the simulated single quantum dot spectra. The parameters we used to match the experimental setup are $r=2.36 nm$ (which is consistent with the manufacturer's estimation of quantum dot size of $r\sim1.5nm$), and a calculationally calibrated ambient force  of $F=1 nN$, corresponding to the an estimation of the force experienced by quantum dots suspend in PMMA.

\section{Conclusions}

A millimeter-scale waveguide made by polymeric materials was manufactured and the core/shell CdSe/ZnS QDs were integrated in the volume of its core.
Since the light source peak (532 nm) is within the wavelength absorption range of the selected QDs, during the transmission through the waveguide the light can be absorbed and eventually re-emitted by the quantum dots at a longer wavelength.

A preliminary experiment was carried out comparing the output spectrum of a non-doped waveguide and the one of a QDs-doped waveguide.
The results showed that a clear distribution stretching on the range [550-600]nm is visible just in the doped waveguide measurement, and it is identifiable as the QDs piezoelectric emission.
Amplitude and wavelength of the light detected at the output of the waveguide are both affected by the pressure applied on the waveguide.
In particular, the calibration experiment showed that the output signal increases in amplitude and blue shifts under compression.
These trends are confirmed by the simulated emission spectra of a single quantum dot under compression shown in Fig.\ref{single}.

Although the QDs optical waveguide can be considered a coupled system of multiple quantum dots (see for example \cite{coupling}), the theoretical model was implemented for a single QD. 
The accordance between experimental and simulation results suggests that the overall phenomenon leads to the same modulation effect of the analysed light properties.
Considering the wavelength range experimentally obtained, it can also be deduced by the simulation analysis that the regime of force sensed by each single QD contained in the waveguide is between [0-0.02]$\mu$N.

A ray-tracing simulation was also carried out to study the behaviour of the multiple quantum dots system of the doped waveguide.
The analysis of the possible events happening within the simulated waveguide core were categorized and the corresponding probability of occurrence were calculated, as shown in Fig.\ref{events}.
A statistical analysis of wavelength distribution and directional information of the simulated detected photons was also presented in Fig.\ref{hist}.

Finally, theoretical amplitude and wavelength shift of the measured photonic distribution measured at the detector as a function of the applied force was obtained in the case of force applied on two different location of the waveguide.
Both in the experimental and simulation results, the trends are more pronounced when the force is applied closer to the detector.
This can be explained by the fact that the closer to the detector this compressive force is applied, the less likely for a QD emitted ray to be scattered out of the waveguide, or absorbed and not re-emitted (these two events would reduce the amplitude variation), or to be absorbed and re-emitted by a subsequent QD (this event would reduce the wavelength variation).
This observation suggests that not only the pressure can be detected, but also the location of the applied force can be obtained, as both light amplitude and wavelength are affected differently through the scattering process.

The accordance between theoretical model and simulation results validates the presented sensing principle.
The calibration experiment demonstrates the feasibility of the proposed force sensor design.
The proposed prototype presents promising capabilities of measuring pressures and detecting their location.
In addition, all the main advantages inherited from the optical sensors potentially make it as a good candidate for several application fields such as nanobiotechnology and robotic sensing.

The main limitation of the prototype is the significant amount of light lost through the waveguide, mainly due to rays being scattered or emitted by quantum dots at angles which do not satisfy the total internal reflection requirements and thereby are transmitted through the cladding and lost.
The efficiency of the QDs waveguide should be further investigated and optimised. 
A study of the QDs concentration or a sensor design improvement (maybe involving additional layers of high refractive index materials which may collect and lead more light to the waveguide output) could help with reducing this drawback.

\begin{acknowledgement}

CW was supported by grants EP/R02992X/1 and EP/R013977/1 from the UK Engineering and Physical Sciences Research Council (EPSRC).

\end{acknowledgement}

\bibliography{citations}

\end{document}